# Highly sensitive magnetic properties and large linear magnetoresistance in antiferromagnetic Cr$_x$Se（0.875≤x≤1）single crystals

Yuqing Bai, Shuang Pan, Ziqian Lu, Yuanyuan Gong, Guizhou Xu[*], Feng Xu[*]

*MIIT Key Laboratory of Advanced Metallic and Intermetallic Materials Technology, School of Materials Science and Engineering, Nanjing University of Science and Technology, Nanjing 210094, China*

**Abstract**

Cr$_x$Se (x≤1) is a class of quasi-layered binary compounds with potential applications in spintronics due to its intriguing antiferromagnetic properties. In this work, Cr$_x$Se single crystals with high Cr content (x=0.87, 0.91 and 0.95) were grown, and their magnetic and transport properties were investigated in detail. It is found that with small increase of Cr content, the Néel temperature ($T_N$) of the samples can dramatically increase from 147 K to 257 K, accompanied with obvious changes in the magnetic anisotropy and hysteresis. The phenomena of field-induced spin-flop transitions were unveiled in these alloys, indicating their comparatively low anisotropy. The magnetoresistance (MR) of the three compounds showed positive dependence at low temperatures and particularly, non-saturated linear positive MR was observed in Cr$_{0.91}$Se and Cr$_{0.95}$Se, with a large value of 16.2% achieved in Cr$_{0.91}$Se (10K, 9T). The calculated Fermi surface and MR showed that the quasi-linear MR is a product of carrier compensation. Our work revealed highly sensitive magnetic and transport properties in the Cr-Se compounds, which can lay foundation when constructing further antiferromagnetic spintronic devices based on them.

**Keywords**: antiferromagnetic alloy; Cr$_x$Se single crystals; spin flop; linear magnetoresistance

---

[*] Corresponding Author: E-mail: gzxu@njust.edu.cn
[*] Corresponding Author: E-mail: xufeng@njust.edu.cn

## 1. Introduction

In recent years, antiferromagnets have attracted increasing interest in the spintronic community due to their technical potential for developing ultra-high-density and high-speed spintronic devices [1,2]. Their intrinsically attractive properties include less susceptibility to external magnetic fields owing to the vanishing net magnetic moment and ultra-fast spin dynamics due to the higher eigenfrequencies[3]. On the other hand, the insensitivity to external field makes the manipulation and detection of antiferromagnetic (AFM) states challenging. An effective way to realize the electrical switching of AFM states is by utilizing the locally asymmetric magnetic sublattices of AFM materials, like in CuMnAs [4] and $Mn_2Au$ [5]. The other way takes advantage of noncollinear AFM materials like $Mn_3GaN$ [6] and $Mn_3Sn$ [7], etc. It is worth noting that in either devices, the read-out of signal is ascribed to the AFM anisotropic magnetoresistance (AMR) or anomalous Hall effect. Therefore, the anisotropic magnetic and transport properties are fundamentally important for the application of AFM materials in spintronic devices.

The chromium-selenium system $Cr_xSe$ ($x \leq 1$) has drawn much attention concerning their intriguing AFM properties. For instance, exchange bias effect due to exchange coupling between the interface spins in AFM and FM has been reported in $Cr_2Se_3$ [8]. Anomalous Hall effect and large negative magnetoresistance were observed in $Cr_{0.68}Se$, which was confirmed to be related to its non-collinear AFM structure [9]. Complex AFM-AFM transition and non-collinear AFM structure were also present below $T_N$ in $Cr_3Se_4$ [10], and spin-flop transition and spin glass state were unveiled in $Cr_7Se_8$ [11]. However, to date, most studies concentrate on the compounds with Cr content lower than 0.87. The possible reason is that the Cr vacancies are always present in this system, making it difficult to synthesize compounds with high Cr content [12]. Nevertheless, by specific method or treatment, $Cr_xSe$ with x approaching 1 can be synthesized. In a remote study [13], $Cr_xSe$ with x=0.93, 0.96 and 1 have been prepared by long-time and high-temperature solid state reactions, where antiferromagnetic interaction is revealed, with $T_N$ varying in the range of 232 K to 279 K. However, in a recent study by Zhang et al. [14], CrSe crystals with grain size reaching the sub-millimeter scale prepared by atmospheric pressure chemical vapor deposition (CVD) were assumed to be ferromagnetic in the ultrathin layer. The contradictory properties are possibly related to the highly sensitive properties of Cr-Se alloys, which deserves to be clarified by more systematic studies.

Therefore, in this work, single crystals of $Cr_xSe$ with high x content (x=0.87, 0.91 and 0.95) were synthesized and their magnetic and transport properties were systematically investigated. It is found that the magnetic properties of the $Cr_xSe$ system are very sensitive to the Cr content. As the Cr content increases with a small amount, the Néel temperature ($T_N$), magnetic anisotropy and hysteresis would change significantly. A field-induced spin-flop behavior was identified in these alloys, which also varied greatly depending on the composition. This highly sensitive and tunable magnetism could provide a promising platform for future device applications. The magnetoresistance (MR) of the three compounds showed positive dependence at low temperatures, which transformed to negative as the temperature increased, and non-saturated linear positive MR was observed in $Cr_{0.91}Se$ and $Cr_{0.95}Se$. The underlying physical mechanisms of linear MR signals were further explored by combining first-principles calculations with Boltzmann transport theory methods.

## 2. Experimental Section

*2.1. Sample preparation and characterization*

Single crystals of $Cr_xSe$ were grown by the flux method. To adjust the Cr content, we have tried different fluxes to synthesize. Firstly, the sample with actual composition of $Cr_{0.87\pm0.07}Se$ was grown with the low melting point metal Ga as the flux. Cr, Se and Ga in the stoichiometric ratio of 2:2:98 was mixed and loaded into the alumina crucible sealed in an evacuated quartz tube. The sealed quartz tube was heated continuously to 1100□ with a duration of 2h, and it was slowly cooled to 730□ at a rate of 3 K/h, afterwards the flux (Ga) was removed by high-speed centrifugation. As a result, air-stable single crystals with a shiny surface and an average size of ~ 4 × 3 × 0.1 mm$^3$ were obtained.

To obtain samples with higher Cr content, Sn was taken as the flux, and the start ratio and the end temperature were also adjusted. When the start ratio of Cr:Se:Sn is 2.3:2:98 and the end temperature is set as 650°C, we can obtain the samples with actual composition of $Cr_{0.91\pm0.10}Se$, while a higher start ratio of 2.4:2:98 can lead to $Cr_{0.95\pm0.08}Se$. But when we continue to increase the start ratio, the Cr content will not increase. The average sizes of the two samples were about $2 \times 2 \times 0.1$ mm$^3$ and $1 \times 1 \times 0.1$ mm$^3$, respectively. All these compositions were examined by using energy-dispersive spectroscopy (EDS) equipped in the scanning electron microscope (SEM, FEI Quanta 250F). To be simplified, Cr$_7$Se$_8$ (for Cr$_{0.87}$Se), Cr$_{0.91}$Se and Cr$_{0.95}$Se were used to describe our samples in the following text.

The structure of samples was identified by x-ray diffraction (XRD, Bruker-AXS D8 Advance) with Cu−Kα radiation. The magnetic and transport properties were characterized by using the Physical Property Measurement System (PPMS, Quantum Design). For measuring the longitudinal resistance, a four-probe method was applied and the electrode contact were made of silver paste. The final MR was symmetrized to exclude the misalignment of the electrode.

*2.2. Calculation details*

First principles calculations were performed with the projector augmented wave (PAW) method, as implemented in the Vienna ab initio simulation package (VASP) [15,16]. The exchange-correlation effect was treated with a generalized gradient approximation (GGA) function in the form of Perdew-Burke-Ernzerhof (PBE) parametrization. The static self-consistency calculations were carried out on a *k* grid of $9 \times 9 \times 9$, with the cutoff energy of 500 eV for the plane wave basis set, and the lattice constants were optimized. To calculate the Fermi surface (FS) and MR, we first construct the maximally-localized Wannier functions (MLWFs) by projecting the Bloch states obtained from VASP onto the Cr 3*d* and Se 3*p* atomic orbitals, using the package of Wannier90 [17]. Then the FS was constructed on a k-mesh of $50 \times 50 \times 50$ using Wannier90 and visualized by Xcrysden [18]. The MR is calculated using the WannierTools [19] software packages with $31 \times 31 \times 31$ *k* points, which is based on the Boltzmann transport theory approach that relies on a semiclassical model and constant relaxation time approximation [20].

3. **Results and discussions**

Fig. 1(a) shows the representative crystal structures of CrSe and Cr$_7$Se$_8$. Based on literatures, ideal CrSe has a typical hexagonal NiAs structure with a space group of P6$_3$/mmc(194), with Cr occupying (0, 0, 0) and Se (2/3, 1/3, 1/4) [9,12]. The X-ray powder diffraction was performed on the crushed single crystals of Cr$_7$Se$_8$, Cr$_{0.91}$Se, and Cr$_{0.95}$Se, along with the CrSe powders prepared by solid-state sintering, as shown in Fig. S1 (supplementary material). It can be seen that for Cr$_{0.91}$Se and Cr$_{0.95}$Se, the supposed hexagonal structure is formed, while distinct peak splitting occurs in Cr$_7$Se$_8$, which corresponds to a monoclinic crystal structure (space group: C2/m) due to a slight lattice distortion, with 4 × 2 × 2 superstructure formed. These findings align with previous studies, when the Cr content is lower than 1:1 and matches the integer atomic ratio, the Cr vacancies tend to arrange in order and are presumed to appear in every second layer of the metal [12,21]. The correlation moments of Cr atoms located on two different layers differ depending on the adjacent environment, which leads to the complexity of the magnetic structure below the $T_N$ [12,22,23]. The fitted lattice parameters are also given on the figure. The XRD patterns of the single crystals are plotted in Fig. 1(b), and all three groups of single crystals show the (00$l$) diffraction peaks of the above crystal structure. With the increase of Cr content, these peaks display a systematic small shift to left, indicating a slight increase of c-axis parameters. Fig. 1(c) shows SEM images and optical photographs of the samples, the prepared single crystals are in millimeter size with quasi-hexagonal planes, in accordance with their crystal structures. Most of them display flat and shining surfaces, but some may have flux residual at the surfaces. These crystals are in thin plate shape, but cannot be mechanically exfoliated, so they are presumed to be quasi-two-dimensional.

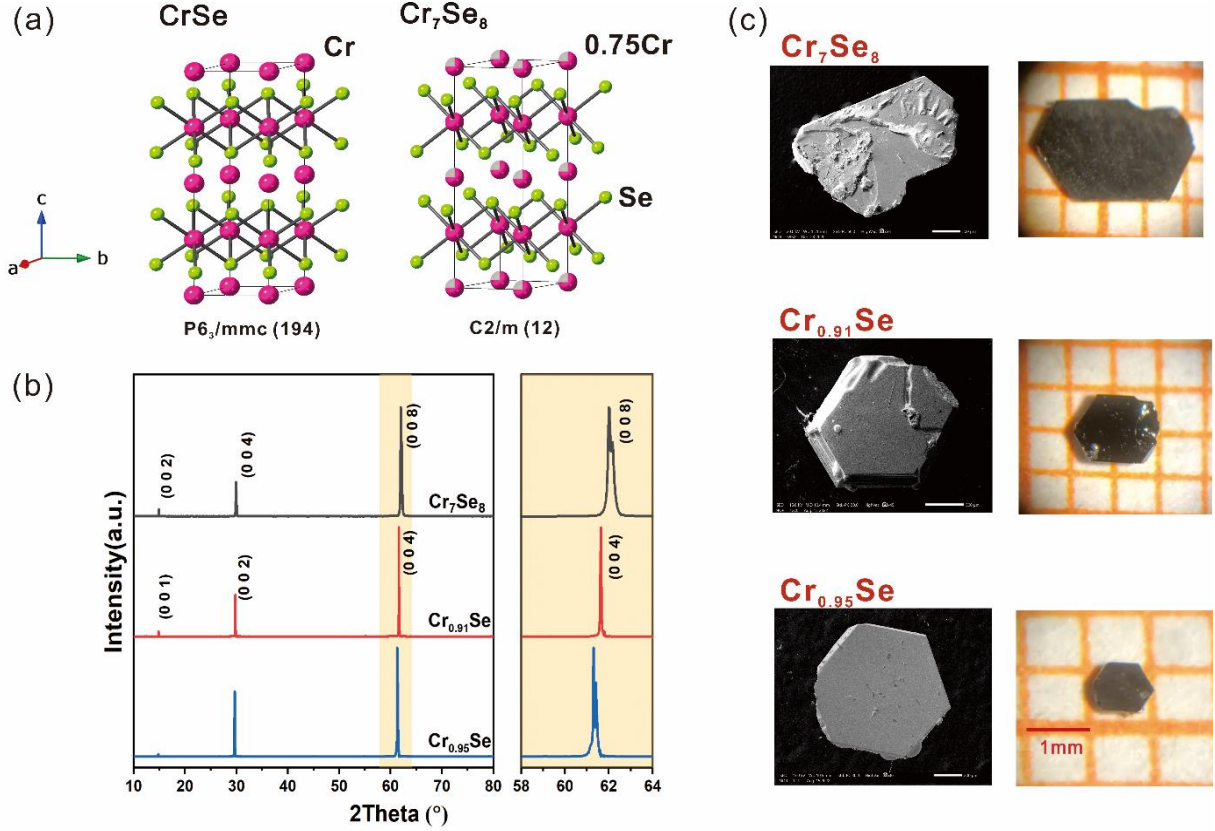

**Fig. 1.** (a) Crystal structures of CrSe and $Cr_7Se_8$; (b) XRD patterns of $Cr_7Se_8$, $Cr_{0.91}Se$ and $Cr_{0.95}Se$ single crystals; (c) SEM images and optical photographs of the samples.

To investigate the magnetic properties of the $Cr_xSe$ single crystals, the magnetization was measured as a function of temperature under a constant field parallel and perpendicular to $c$ axis, respectively. Fig. 2(a)-(c) show dc susceptibilities ($\chi=\frac{M}{H}$) vs $T$ curves under an applied field of $H$=1kOe. It can be seen that the transition temperatures significantly vary with small changes of Cr content. For $Cr_7Se_8$, the AFM transition takes place at $T_N$ ~147K, and the divergence of the susceptibility is small for the two directions, indicating small magnetic anisotropy of it. For $Cr_{0.91}Se$, the susceptibility for $H//ab$ exhibits a sharp peak at 198 K, indicative of the AFM transition at this temperature ($T_N$). While for $H//c$, only a slight kink is found at $T_N$. In addition, below $T_N$, $\chi(T)$ curves of $Cr_{0.91}Se$ show another obvious magnetic transition at ~81K (denoted as $T_t$), which is supposed to be a transition between two different AFM states, implying the complexity of the magnetic structure in this compound, as also observed in other AFM materials [10,24,25]. For $Cr_{0.95}Se$, only a typical AFM transition at $T_N$ = 257K is observed along $H//ab$, while the susceptibility increases along *H//c*, approaching the paramagnetic (PM) behavior. It is due to the fact that the magnetic moment mainly orients in the *ab* plane, as also observed in some other AFM systems[9,26,27]. The PM to AFM transition temperature $T_N$ and AFM-AFM transition temperature $T_t$ of these three alloys are plotted in Fig. 2(d), along with the magnetic transition temperatures of some other chromium-selenium compounds that have been reported [9,11,13,28,29]. It can be seen that, as the Cr content x increases, the $T_N$ presents a monotonically growing trend, and climbs more quickly at high Cr content. This reflects the sensitivity of the exchange coupling to the composition in chromium-selenium compounds, which might be caused by the subtle competition of the direct Cr-Cr exchange interaction and super-exchange interaction between Cr-Se-Cr [10,30–32].

The $T_t$ transition in $Cr_{0.91}Se$ was further inspected by the in-plane ac susceptibility measurement under constant external field (*H*=1000 Oe) and in the temperature range of 10–300 K (Inset of Fig. 2(b)). The transitions at $T_t$ and $T_N$ can be well resolved in the real component *χ'* (left axis). Furthermore, an additional peak at ~40K appears in both the *χ'(T)* and *χ''(T)* (right axis) curves, which is defined as $T_F$ (shown by arrow). The anomaly at $T_F$ is thought to be related to the weak ferromagnetism in the low temperature region [25], as also reflected in the magnified in-plane *χ(T)* curves. The possible reason is that, the non-stoichiometry of our samples can lead to disorder of the Cr vacancies, thus induce local ferromagnetic clusters [11,12]. These clusters, which may be spin glass or other magnetic states, result in weak ferromagnetism in the low-temperature region.

In order to further understand the magnetic behavior of the three alloys, the Curie-Weiss law was applied to fit the *χ(T)* curves (along *H//ab*) above the $T_N$, as shown by the black solid lines in Figs. 2(a)-(c). The temperature dependences of the inverse susceptibility (1/*χ*) in both crystallographic directions are provided in Fig. S2 (supplementary material), where linear relationships above the Néel point are observed, indicating good accordance to the Curie-Weiss behavior. For $Cr_{0.95}Se$, the fit is also performed between 48K and 257K (along *H//c*), it starts to deviate severely from the original curve when the temperature is lower, indicating the emergence of a weak ferromagnetism. The Curie-Weiss law is described as [33]

$$\chi(T) = \frac{M}{H} = \frac{C}{T-\theta_P} + \chi_0 \qquad (1)$$

where $C$ is the Curie constant, $\theta_P$ is the Weiss temperature, and $\chi_0$ is the Pauli PM constant. The fitted parameters and their corresponding errors for the three samples are given in Table 1. The negative $\theta_P$ are obtained for all the samples, the values of $|\theta_P|$ become larger with increase of x, in line with their AFM ordering and the variation of $T_N$.

The effective magnetic moment $\mu_{eff}$ are also obtained by

$$\mu_{eff}/\mu_B = \sqrt{3k_B C/N_A Z}$$

(2)

where $k_B = 1.38 \times 10^{-16}$ erg/K, $N_A = 6.02 \times 10^{23}$ mol$^{-1}$, and $Z$ is the atom number per unit cell. The $\mu_{eff}$ of the three samples are listed in the Table 1, which are close to their theoretical values when considering the spin-only moments of $Cr^{2+}$ (S=2) and $Cr^{3+}$ (S=3/2), which are $4.9\mu_B$ and $3.87\mu_B$, respectively. The ratios of $Cr^{2+}$ and $Cr^{3+}$ in different samples are determined according to the principle of electric neutrality. (See supplementary material for calculation details)

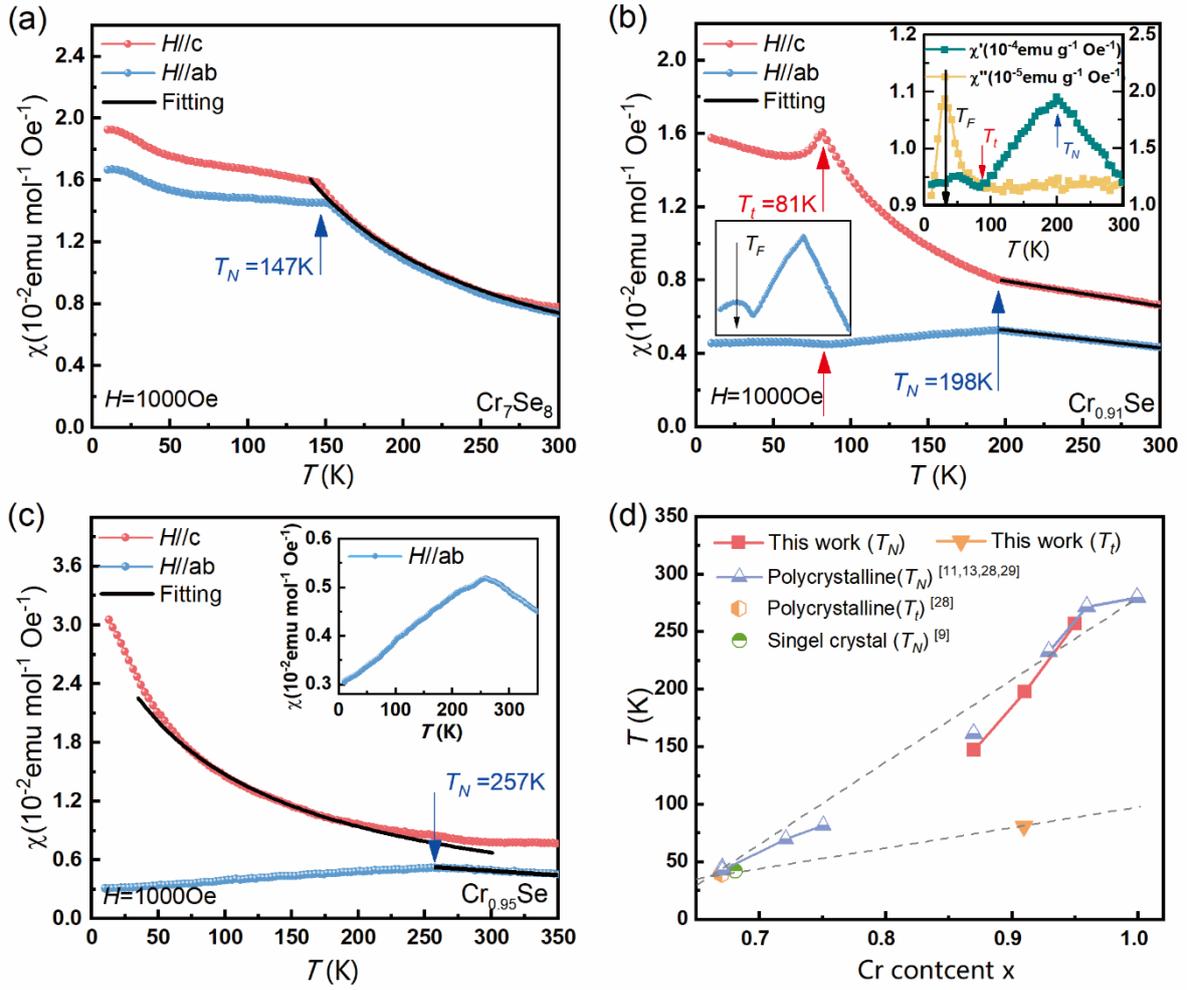

**Fig. 2.** (a)-(c) Temperature dependences of magnetization along $H//c$ and $H//ab$ at $H = 1000$ Oe for all samples; The black solid lines indicate the fitting results according to the Curie-Weiss law. The left inset in (b) shows the enlarged $M(T)$ curve for $H//ab$, the right inset shows the temperature dependences of ac susceptibility at constant field ($H$=1000 Oe) and frequency ($f$ = 99 Hz); Inset in (c) shows the enlarged $M(T)$ curve for H$//ab$; (d) Comparison of $T_N$ and $T_t$ measured in this and other works.

Table 1 The fitted parameters and errors of Curie-Weiss law for $Cr_7Se_8$, $Cr_{0.91}Se$ and $Cr_{0.95}Se$, and the calculated effective moments of the Cr ions.

| Samples | | C (emu K/mol) | $\theta_p$ (K) | $\chi_0 (\times 10^{-5})$ | $\mu_{eff}$ ($\mu_B$) | Cal. $\mu_{eff}$ ($\mu_B$) |
|---|---|---|---|---|---|---|
| $Cr_7Se_8$ | $H//ab$ | 2.11±0.12 | -1.68±0.76 | -2.08±0.62 | 4.11±0.11 | 4.61 |
|  | $H//c$ | 2.13±0.19 | -3.09±1.23 | 13.4±3.18 | 4.13±0.18 |  |
| $Cr_{0.91}Se$ | $H//ab$ | 2.74±0.02 | -272.78±5.87 | -1.45±0.55 | 4.68±0.02 | 4.70 |
|  | $H//c$ | 2.72±0.13 | -203.53±6.12 | 2.40±0.98 | 4.67±0.10 |  |
| $Cr_{0.95}Se$ | $H//ab$ | 2.93±0.06 | -372.62±9.04 | -1.72±0.47 | 4.84±0.05 | 4.79 |
|  | $H//c$ | 2.92±0.08 | 318.80±7.23 | 2.35±1.13 | 4.83±0.07 |  |

In the measurement of the susceptibility, it is noticed that the magnetic anisotropy also varies with x. Hence the isothermal magnetization $M(H)$ curves are further investigated at various temperatures between 10 and 100 K for both $H//ab$ and $H//c$ directions. To measure the $M(H)$ curves, the samples were first cooled to 10K under the measurement field of $M(T)$ curves (1kOe), then warmed to each measurement temperature at zero field. The field sequence is 0T-9T-0T, including the virgin curve. There are several features that are noteworthy: i) Magnetic field-induced magnetization jumps were observed in a certain temperature range below $T_N$ in all the three samples, almost along both directions, except for $Cr_{0.95}Se$. ii) Large magnetic hysteresis exists in $Cr_7Se_8$, while it decreases in $Cr_{0.91}Se$ and totally vanishes in $Cr_{0.95}Se$. iii) For $Cr_7Se_8$, the $M(H)$ behaviors along the two directions are nearly superposed, like the case of $Cr_{0.68}Se$ [9], implying the approximate isotropic magnetic properties of it. While the magnetic anisotropy becomes evident with increasing Cr content. At x=0.95, the magnetization for $H//c$ grows sub-linearly with external field, while that in the $ab$ plane still undergoes steep jumps.

These discontinuous magnetization jumps are the so-called field-induced spin-flop (SF) transitions, which can take place in collinear AFM systems with weak magnetic anisotropy [34–37]. This is because despite the magnetic moments are preferentially aligned to the easy axis due to the magnetic anisotropy, they will reorient themselves and form a canted configuration perpendicular to the easy axis at a critical field (schematically shown in the inset of Fig. 3(c)), thus reducing the magnetic energy by $1/2(\chi_\perp - \chi_{//})H^2$, $\chi_\perp$ and $\chi_{//}$ are the susceptibilities perpendicular and parallel to the easy axis, respectively [36,38]. The critical SF field, marked as $H_{SF}$, is scaled by $(H_{SF})^2 = 2K/(\chi_\perp - \chi_{//})$, where $K$ is the anisotropy constant [36,39]. In the three compounds concerned here, the $M(H)$ curves of $Cr_7Se_8$ possess the smallest anisotropy, hence the $H_{SF}$ is also the lowest among them. The large hysteresis in it is supposed to associate with the existence of ferromagnetic-like clusters in the AFM matrix [11]. While in $Cr_{0.95}Se$, the SF transition is observed only for $H//ab$, indicating the easy axis deflects to in-plane, like the case of $NiPS_3$ [36]. Overall, the $H_{SF}$ tends to increase with high temperatures, due to the decreasing difference between the $\chi_\perp$ and $\chi_{//}$.

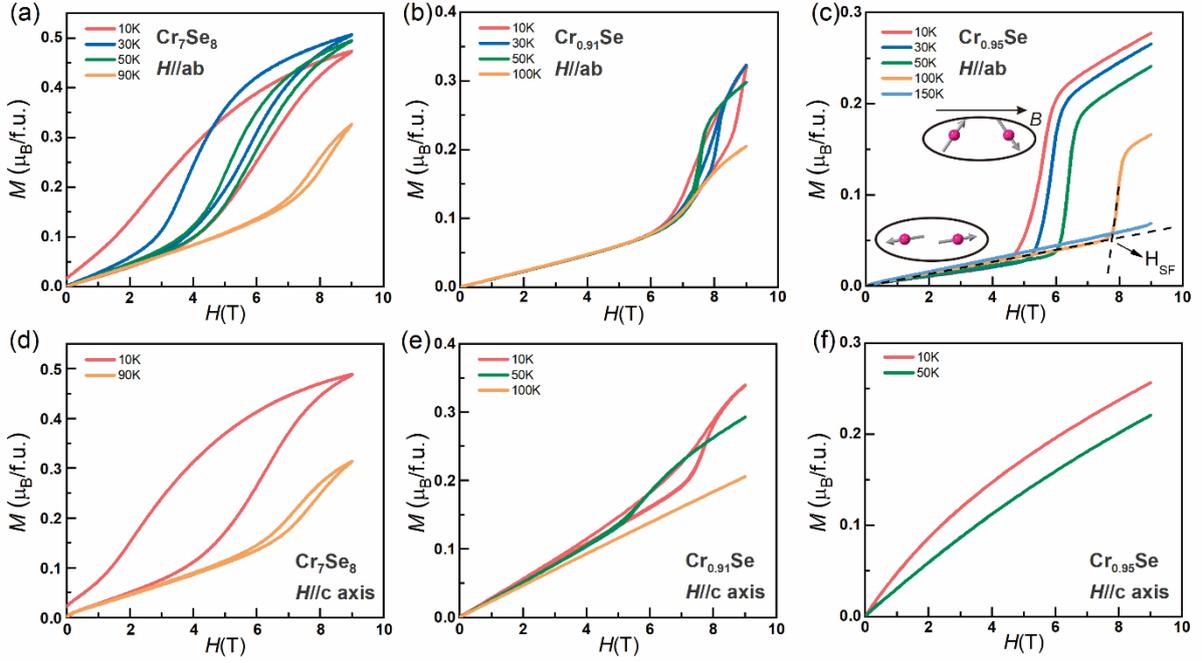

**Fig. 3.** Isothermal $M(H)$ curves of $Cr_7Se_8$, $Cr_{0.91}Se$ and $Cr_{0.95}Se$ from 10 K-150K for $H//ab$ (a-c) and $H//c$ (d-f). Inset in (c) is a schematic diagram of SF transition.

In the following, the magnetoresistance (MR = $\frac{\rho(B)-\rho(0)}{\rho(0)}$) for $Cr_{0.91}Se$ and $Cr_{0.95}Se$ are investigated with $H \perp I$ (i.e. $H//c$) and $H//I$ ($I$ lies in the $ab$ plane). For $Cr_7Se_8$, as the $M-H$ behaviors are isotropic along the perpendicular and parallel directions, the MR data of both directions are also nearly superposed, so only one set of data ($H \perp I$) is present. The MR curves were obtained by sweeping the magnetic field from 9T to -9T, and for $Cr_7Se_8$, back to 9T again. The samples were cooled to the target temperature under the stopping field of the previous measurement (9T or -9T). For clarity, all the MR curves are shifted vertically. As shown in Figs. 4(a-c), the three compounds share a common feature of that, the MR is positive at low temperatures, and turns to negative with higher temperatures (above 30 or 50K). At relatively high temperatures, the suppression of spin disorder scattering by the applied magnetic field dominates, resulting in negative MR. This contribution starts to diminish below a certain temperature and the MR due to Lorentz force starts to become significant. The MR behavior of $Cr_7Se_8$ at 50 K [Fig. 4(a)] shows a clear competition between the two effects, and similar behaviors have been found in other antiferromagnetic materials [24,41]. What is particularly noteworthy here is the large positive and almost linear MR at low temperatures, especially for $H\perp I$ in $Cr_{0.91}Se$, which reaches about 16.2% at 9T, higher than most of the widely concerned AFM metal materials [40,41]. It is known that the dependence of the Lorentz MR on the magnetic field is usually parabolic at low magnetic fields, unlike the quasi-linear relationship here. The reason for this will be discussed later by further considering the specific FS of CrSe.

However, the details of the MR are largely varied with different compositions, which are related to their distinct magnetization behaviors. In $Cr_7Se_8$, resembling to its *M(H)* curves, large hysteresis is also observed in MR. However, due to the different field sequence of the MR and MH, some details, like the closing field at 10K, might not completely be the same. In addition, the MR curves will usually turn at the SF field, which are most pronounced in $Cr_7Se_8$ and *H//ab* of $Cr_{0.95}Se$. The MR turn can be attributed to the different ordering of spin structure across the SF transition. Meanwhile, changes in the electronic structures can also have an impact on the MR. For example, the emergence of superzone band gaps[42,43], resulting from the enlarged unit cell of the noncollinear antiferromagnetic structure, can be one possible reason for the increase in resistivity, leading to the observed MR jump.

The temperature dependences of the resistivity ($\rho(T)$) for the three samples are also shown in Fig. 4(d), which exhibit clear metallic behavior. When come across the Néel transition, a faint but discernable kink can be observed in the curves, which is more clearly seen in the derivative of $\rho$ (d$\rho$/d$T$).

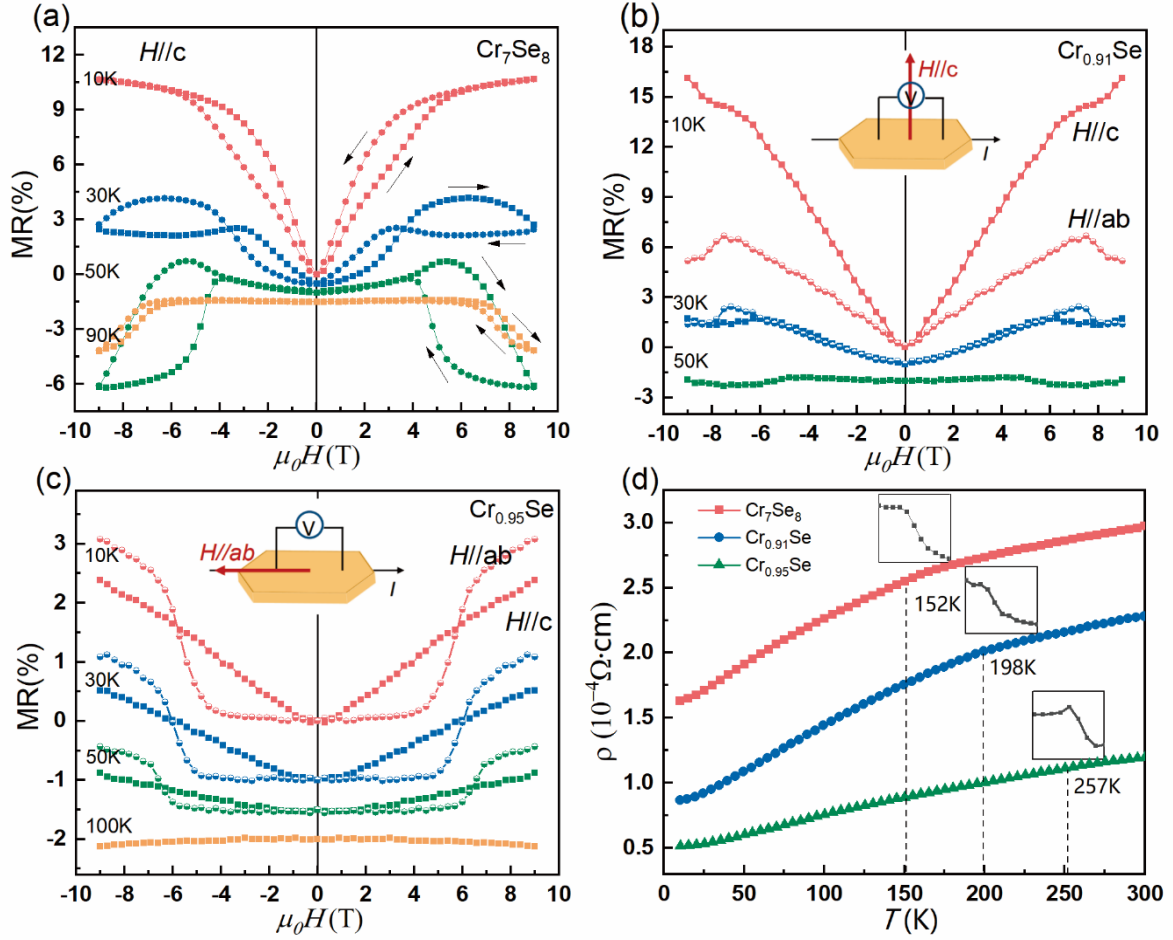

**Fig. 4.** (a)-(c) Field dependences of the MR of $Cr_7Se_8$, $Cr_{0.91}Se$ and $Cr_{0.95}Se$ at various temperatures for $H//c$ ($H \perp I$) and $H//ab$ ($H // I$). The MR curves are shifted vertically for clarity. Insets are the schematic of the measurement configuration for (b) $H//c$ and (c) $H//ab$; (d) Temperature dependences of the resistivity of $Cr_7Se_8$, $Cr_{0.91}Se$ and $Cr_{0.95}Se$. Insets show the derivative of ρ (dρ/d$T$) near the transition temperature.

To gain a deeper understanding of the field dependence of the positive linear MR in the low temperature region, the magnetoresistance of CrSe has been directly calculated using the packages of Wannier90 and WannierTools (see the methods).

Fig. 5 shows the band structure and Fermi surface of CrSe in the AFM state. It can be seen that only two bands pass through the Fermi energy ($E_F$), which is marked as the 10th and 11th band. They touch each other at several high symmetry points (K, L, H), which leads to a low density of states at $E_F$. These two bands result in the two Fermi surfaces shown in Figs. 5(b, c). The one formed by the 10th band contains a large distorted sphere centered at the A high-symmetry point, which corresponds to the band that crosses the $E_F$ along the Γ−A and A-L directions. It also includes smaller and discrete triangular structures around the H and L points. The 11$^{th}$ band consists of a closed central pocket (Γ point) and a large multiply connected surface (Fig. 5(c)). As the red side represents the occupied state, so they are all electron orbitals. The FS in Fig. 5(b), on the other hand, corresponds to the hole orbitals, because the blue side is the occupied state and there are no electrons in the region enclosed by the FS. Therefore, carrier compensation exists in this system that can affect the transport properties.

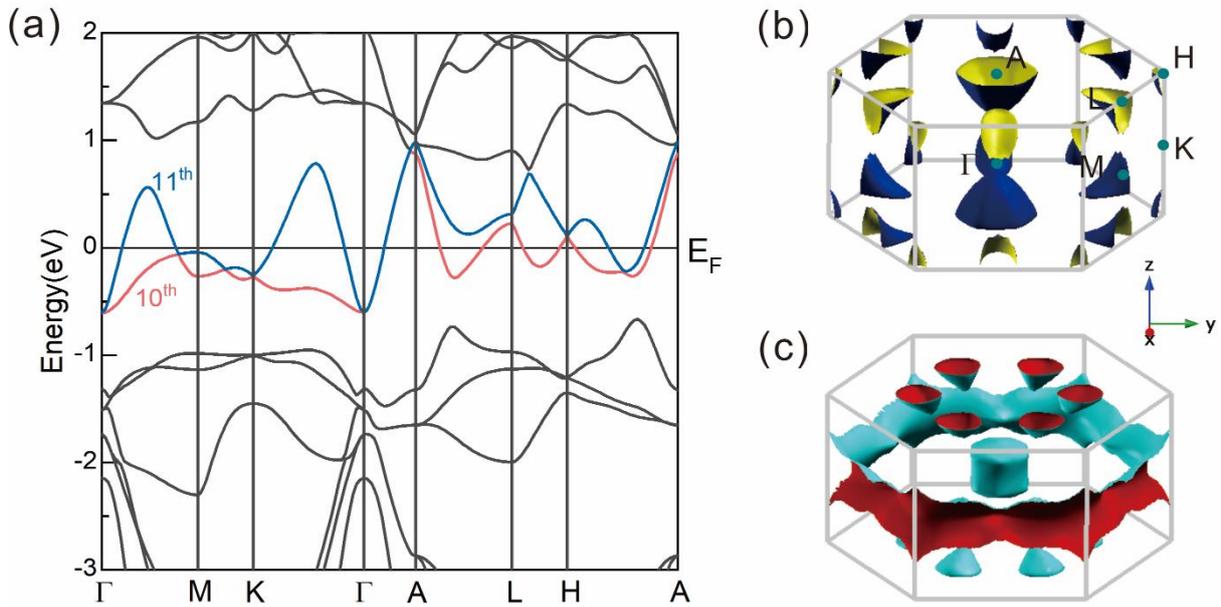

**Fig. 5.** (a) Band structure of antiferromagnetic CrSe. (b)-(c) Fermi surface of CrSe arising from (b)10$^{th}$ (c) 11$^{th}$ bands. The high symmetry positions in the first Brillouin zone (BZ) are Γ (0, 0, 0), M (1/2, 0, 0), K (1/3, 1/3, 0), A (0, 0, 1/2), L (1/2, 0, 1/2) and H (1/3, 1/3, 1/2), respectively.

The resistivity tensor (by inverted of the conductivity σ) of CrSe is directly calculated in presence of an applied magnetic field for different directions and temperatures. The current direction is fixed along the *y* axis, that is the crystallographic *b* axis, hence the longitudinal resistivity is represented by $\hat{\rho}_{yy}$, as manifested in Fig. 6. As the relaxation time τ is an uncertain factor, both of the resistivity and magnetic field are revealed as a combined variable by multiplying τ. It is found that when *H*//*z* (*c* axis), the MR is essentially linear without tendency of saturation at 10 K, and gradually deviates from a straight line with increasing temperature, in highly accordance with that observed in $Cr_{0.91}Se$ and $Cr_{0.95}Se$. Since in this direction, the electron and hole orbits are both present, and conduction electrons all undergo cyclotron motion along the closed orbitals, the carrier compensation is thought the main reason that caused the quasi-linear MR here. Similar phenomenon has been widely observed in the electron-hole compensated semimetals [44,45]. For *H//y* (*H//I*), MR also increases linearly in the low field region, but slows down in the high field region. Similar to the situation observed in some metals [46,47], the MR tends to saturate at high magnetic fields due to incomplete compensation of the two kinds of charge carriers in specific magnetic field directions. Obviously, the compensation in the *y* direction is weaker than in the *z* direction. The MR turn across the SF transition is not accounted in the calculation, since the exact spin structure is not very clear here.

Finally, it is observed that the sample $Cr_{0.91}Se$ exhibits the largest MR, which cannot be explained by the perfect structure model here. Since in real metals, there may be structure defects, like vacancies, disorder or impurities, etc. Here when the Cr concentration deviates from the proportional value (x=7/8 or 1), the crystal defect is supposed to be maximized, which can heighten the spin disorder scattering , leading to large MR [48].

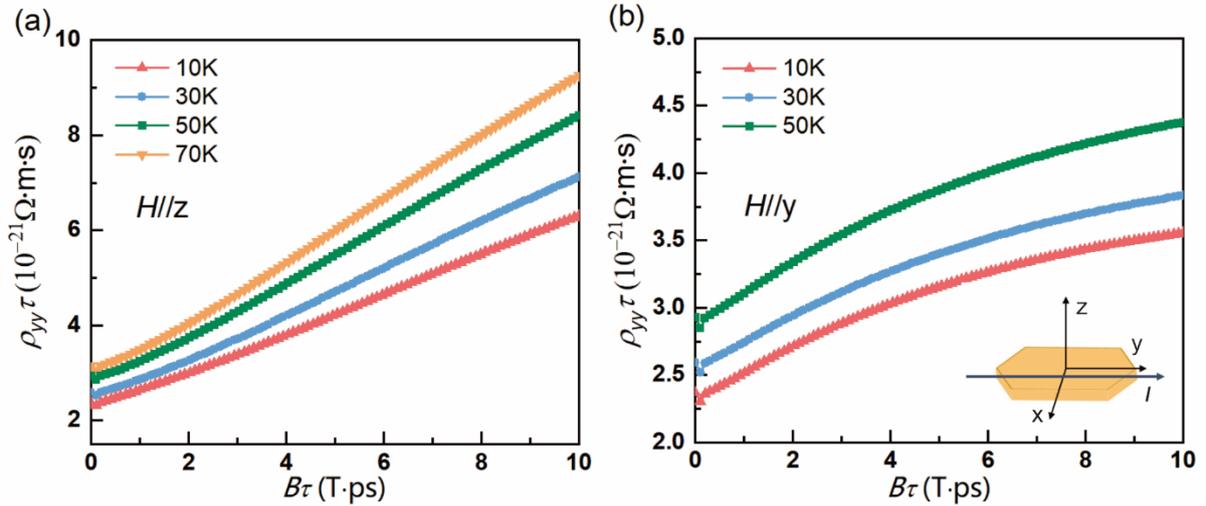

**Fig. 6.** Calculated magnetoresistance at various temperatures for CrSe with magnetic field along the (a) z and (b) y directions.

## 4. Conclusion

In summary, the magnetic and transport properties of the three hexagonal chromium-selenium compounds, $Cr_7Se_8$, $Cr_{0.91}Se$ and $Cr_{0.95}Se$, are investigated after successfully synthesizing their single crystals. It is found that with small increase of Cr content, the $T_N$ of the samples can dramatically increase from 147 K to 257 K, accompanied with obvious change in the magnetic anisotropy and hysteresis. The $Cr_7Se_8$ single crystal exhibits almost isotropic electrical and magnetic behaviors, while the $Cr_{0.95}Se$ presents obvious anisotropy, which is originated from the gradual deflection of the spin orientation from the canted state to in-plane. Hysteresis phenomenon is detected in $Cr_7Se_8$ and the low temperature region of $Cr_{0.91}Se$, owing to the possible ferromagnetic clusters existed in the AFM matrix. Furthermore, field-induced SF transitions are identified in all samples and are very sensitive to the composition. Uncovering of this highly sensitive magnetism can help address some controversial issues regarding this system, and provide guidelines when trying to take advantage of its antiferromagnetic properties.

The MR signals of the three compounds showed positive dependence at low temperatures, which transform to negative as the temperature increases, as a result of the competition between the electron orbital scattering by the Lorentz effect and spin disorder scattering. Remarkably, non-saturated linear positive MR is observed in $Cr_{0.91}Se$ and $Cr_{0.95}Se$. By calculating the FS acquired from first principles methods and MR based on Boltzmann transport theory, it is suggested that the quasi-linear MR is mainly a product of carrier compensation.

## Acknowledgements

This work is supported by National Natural Science Foundation of China (Grant No. 11604148).